\documentclass[]{epl2}

\usepackage{amssymb}
\usepackage{amsmath}
                                                         
\def\bsigma{\mbox{\boldmath $\sigma$}}

\def\bxi{\mbox{\boldmath $\xi$}}

\def\tensor{\overset{\leftrightarrow}} 

\title{Mesoscopic model for the fluctuating \\ hydrodynamics of binary and 
ternary mixtures}
\shorttitle{Mesoscopic model for binary and ternary mixtures}
 
\author{E. T{\"u}zel, G. Pan, T. Ihle \and 
D.M. Kroll}      
\shortauthor{E. T{\"u}zel et al.}    
\institute{
 Department of Physics, North Dakota State University, P.O. Box 5566,
Fargo, ND 58105, USA
}

 
\abstract{     
A recently introduced particle-based model for fluid dynamics with 
continuous velocities is generalized to model immiscible binary 
mixtures. Excluded volume interactions between the two components  
are modeled by stochastic multiparticle collisions which depend on 
the local velocities and densities. Momentum and energy are conserved 
locally, and entropically driven phase separation occurs for 
high collision rates. An explicit expression for the equation 
of state is derived, and the concentration dependence of the bulk free 
energy is shown to be the same as that of the Widom-Rowlinson model. 
Analytic results for the phase diagram are in excellent agreement 
with simulation data. Results for the line tension obtained from the 
analysis of the capillary wave spectrum of a droplet agree with 
measurements based on the Laplace's equation. The introduction of 
``amphiphilic'' dimers makes it possible to model the phase behavior 
and dynamics of ternary surfactant mixtures.}   

\begin{document} 

\maketitle 
 
\section{Introduction}

Hydrodynamic interactions and thermal fluctuations play a crucial role 
in a wide range of phenomena in soft matter physics and molecular and 
cellular biology. Because of the complexity of these systems, simulations 
have played an essential role in much of the research in these areas. In 
fact, the wide range of length and time scales in these problems places 
severe requirements on simulation protocol, and has lead to the 
development of several new 
coarse-grained, mesoscale simulation techniques such as lattice gas 
automata \cite{frisch_86_lan}, the lattice Boltzmann method 
\cite{succi_01_lbe}, dissipative particle dynamics 
\cite{hoogerbrugge_92_smhpwdpd,groot_97_dpdbbams}, smoothed particle 
hydrodynamics \cite{espanol_03_sdp}, and a newer approach variously called 
multi-particle collision dynamics or stochastic rotation dynamics (SRD) 
\cite{malevanets_99_mms}. 
The basic motivation of all these approaches is to coarse-grain out 
irrelevant atomistic details while correctly incorporating 
the essential physics and conservation laws. 

SRD has several attractive features which have lead to its use in studies  
ranging from sedimentation in colloidal suspensions \cite{padding_04_hbf} 
to the dynamic behavior of polymers in solution \cite{falck_03_dst,mussawisade_05_dpp} and vesicles in 
flow \cite{noguchi_04_fvw}.  
In particular, it enables simulations in the microcanonical ensemble 
while fully incorporating both hydrodynamic interactions and thermal 
fluctuations; in addition, because SRD is a particle-based method, the 
coupling to colloidal particles, polymers, or other aggregates is 
straightforward, and the Brownian motion of these embedded objects is 
realized in a very natural way---through random collisions with the solvent 
particles. Finally, the simplicity of the algorithm has made it possible 
to obtain accurate analytic expressions for the transport coefficients~\cite{ihle_05_ect,pooley_05_ktd,tuzel_06_dcs}. 

Recently, it has been shown how to generalize the multi-particle collisions 
of the original SRD algorithm to model excluded volume effects, allowing for a 
more realistic modeling of dense gases and liquids with a nonideal equation 
of state~\cite{ihle_06_cpa,tuzel_06_ctc}. 
In this Letter we show that a similar approach can be used  
to model immiscible binary mixtures. The resulting model retains much of the 
simplicity of the original SRD algorithm, allowing for a detailed analysis
of the transport coefficients and an explicit calculation of the 
equation of state and the bulk entropy and free energy densities. Since there 
is no potential energy, all interactions in the model are entropic, and 
the resulting bulk free energy density is the same as that of the 
Widom-Rowlinson model \cite{rowlinson_02_mtc}. Theoretical 
predictions for the phase diagram are shown to be in good agreement with 
simulation data, and results for the line tension obtained from 
an analysis of the spectrum of capillary wave fluctuations  
agree with measurements based on the Laplace equation. 

\section{Model}  

In a binary mixture of A and B particles, phase separation can occur 
when there is an effective repulsion between A-B pairs. In the current 
model, this is achieved by introducing multi-particle collisions between 
A and B particles. The fluid is modeled by a large number $N$ of point-like 
particles of unit mass which move in continuous space with
a continuous distribution of velocities. There are $N_A$ and $N_B$ particles 
of type A and B, respectively. In two dimensions, the system is 
coarse-grained into $(L/a)^2$ cells of a square lattice of linear dimension 
$L$ and lattice constant $a$. The generalization to three dimensions 
is straightforward. 

\begin{figure}
\twofigures[height=2.5in]{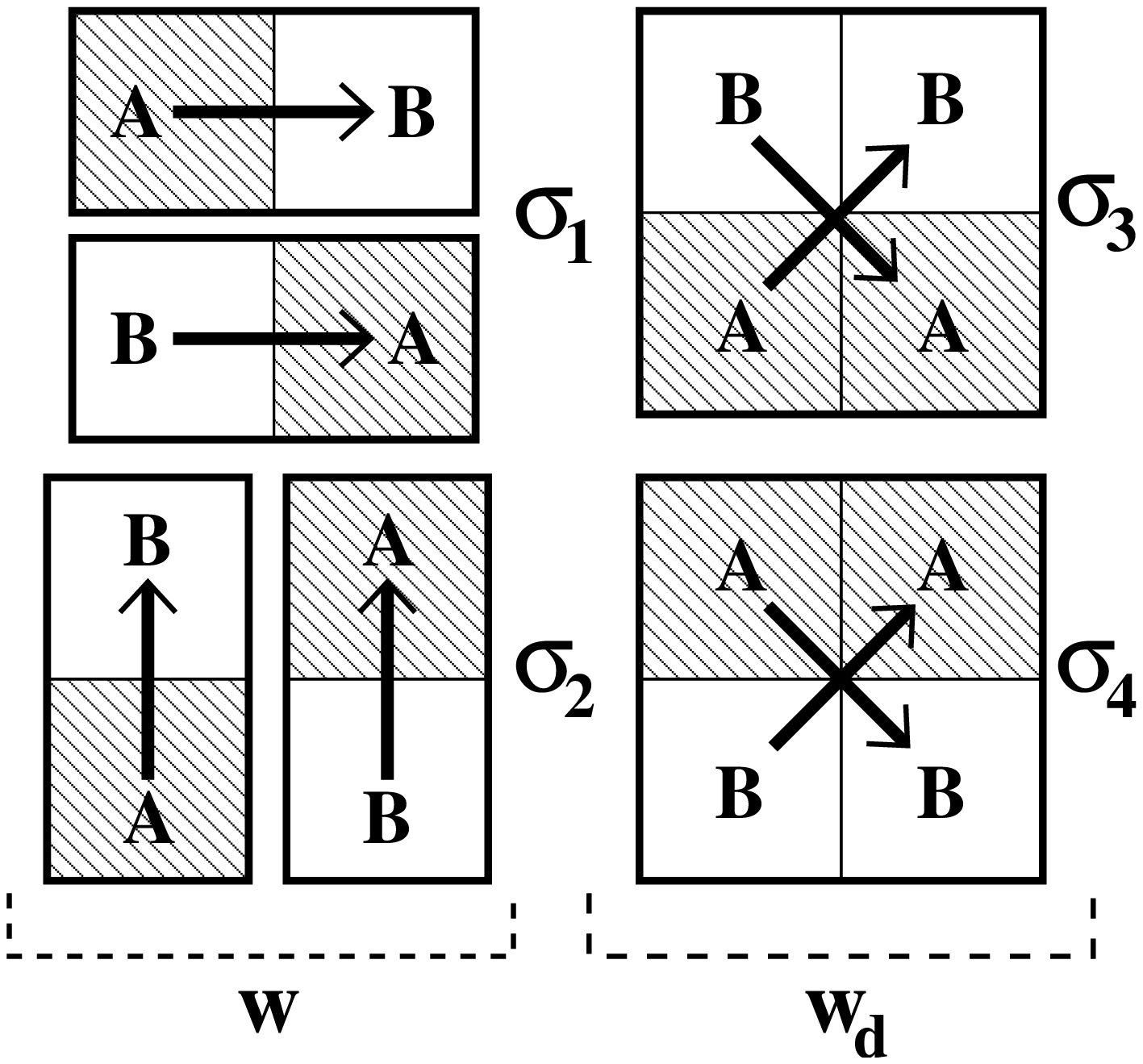}{fig2.eps}
\caption{Schematic binary collision rules. Momentum is exchanged in three ways: a) horizontally along $\bsigma_1$, b) vertically along $\bsigma_2$, and c) diagonally along $\bsigma_3$ and $\bsigma_4$. $w$ and $w_d$ denote the probabilities of choosing collisions a), b) and c), respectively. For a cell pair defined by the vector $\bsigma$, 
A particles in one cell collide with B particles in the other cell, and vice versa.}
\label{fig.1}
\caption{Non-ideal contribution to the pressure, $P_n$, (in units of $k_BT/a^2$) as a function of the tuning parameter $A$ (in units of $\tau/a$). The bullets ($\bullet$) and open circles ($\circ$) are results for 
$k_BT=0.006$ and $k_BT=0.003$, respectively. The lines are plots of Eq. (\ref{PTH}). The inset shows the pressure difference across a droplet interface as a function of the 
inverse droplet radius, for $A=0.60$ at $k_BT=0.0005$. The pressure is measured using the expression given in Eq. (\ref{VSP}). The solid line is a plot of Eq. (\ref{LE}), using the line tension obtained from the analysis of capillary waves.
Parameters: $L/a=64$, $M_A \equiv  N_A / (L/a)^2 = 5$, 
$M_B \equiv N_B / (L/a)^2 = 5$, $a=1$, and $\tau=1.0$.}
\label{fig.2}
\end{figure}

To define the collisions, we introduce a second grid with sides of length 
$2a$ which groups four adjacent cells into one ``supercell''. As discussed 
in \cite{ihle_01_srd}, Galilean invariance requires that the collisions occur 
in a randomly shifted grid. All particles are shifted by the {\it same}
random vector with components in the interval $[-a,a]$ before the collision
step.  Particles are then shifted back by the same amount
after the collision. To initiate a collision, pairs of cells in every supercell
are selected at random. As shown in Fig. \ref{fig.1}, three different choices 
are possible: a) horizontal (with $\bsigma_1=\hat x$), b) vertical ($\bsigma_2
=\hat y$), and c) diagonal collisions (with $\bsigma_3=
(\hat x+\hat y)/\sqrt{2}$ and $\bsigma_4=(\hat x-\hat y)/\sqrt{2}$). 
For each pair of cells, two types of collisions are possible. As illustrated 
in Fig. \ref{fig.1}, particles of 
type A in the first cell can undergo a collision with particles of type B 
in the second cell; vice versa, particles of type B in the first cell can 
undergo a collision with particles of type A in the second cell. There are 
no A-A or B-B collisions, so that there is an effective repulsion between 
A-B pairs. The rules and probabilities for these collisions are chosen in 
the same way as in the nonideal single-component fluid described in 
Refs. \cite{ihle_06_cpa,ihle_06_sdp}. For example, consider the collision 
of A particles in the first cell with the B particles in the second.  
The mean particle velocity of A particles in the first cell is 
${\bf u}_A=(1/m_A)\,\sum_{i=1}^{m_A}\,{\bf v}_i$,
where the sum runs over all A particles, $m_A$, in the first cell.
Similarly, ${\bf u}_B=(1/m_B)\,\sum_{i=1}^{m_B}\,{\bf v}_i$ is the mean 
velocity of B particles in the second cell. The projection of the 
difference of the mean velocities of the selected cell-pairs on 
${\bf \sigma}_j$, $\Delta u_{AB}={\bf \sigma}_j\cdot ({\bf u}_A-{\bf u}_B)$,
is then used to determine the probability of collision.
If $\Delta u_{AB}<0$, no collision will be performed. For positive 
$\Delta u_{AB}$, a collision will occur with an acceptance probability 
\begin{equation}
\label{NONID0}
p_A(m_A,m_B,\Delta u_{AB})={\rm max} \{1, A\,m_A m_B \,\Delta u_{AB}\,\Theta(\Delta u_{AB})\}\,  ,
\end{equation}
where $\Theta$ is the unit step function and $A$ is a parameter which
allows us to tune the equation of state; in order to ensure thermodynamic 
consistency, it must be sufficiently small that $A\,m_A m_B \,\Delta u_{AB}\,\Theta(\Delta u_{AB})<1$ for essentially all collisions~\cite{tuzel_06_ctc}. When a collision occurs, 
the parallel component of the mean
velocities of colliding particles in the two cells is exchanged, 
$v_i^{\Vert}(t+\tau)-u_{AB}^{\Vert}=-(v_i^{\Vert}(t)-u_{AB}^{\Vert}) $,
where  $u_{AB}^{\Vert}=(m_Au^{\Vert}_A+m_Bu^{\Vert}_B)/(m_A+m_B)$ is 
the parallel component of the mean velocity of the colliding 
particles. The perpendicular component remains
unchanged. It is easy to verify that these rules conserve momentum and
energy in the cell pairs. The collision of B particles in the 
first cell with A particles in the second is handled in a similar 
fashion. 

Because there are no A-A and B-B collisions, additional SRD collisions 
at the cell level are incorporated in order to mix particle momenta. 
The order of A-B and SRD collision is random, 
i.e., the SRD collision is performed first with a probability of one half. 
If necessary, the viscosity can be tuned by not performing 
SRD collisions every time step. The results presented in this Letter 
were obtained using a SRD collision angle of $90^\circ$. 

\section{Transport Coefficients} 

The transport coefficients can be determined using the same
Green-Kubo formalism as for the original SRD algorithm 
\cite{ihle_05_ect,ihle_01_srd,ihle_03_srd_a} and its extensions 
\cite{ihle_06_cpa,tuzel_06_ctc,ihle_06_sdp}. In particular, the 
discrete Green-Kubo relation 
\begin{equation}
\label{VTCGK}
\Lambda_{\alpha\beta}(\hat{\bf k})  = 
\frac{\tau}{Nk_BT}\left.\sum_{n=0}^\infty\right.'
g_{\alpha\beta}(\hat{\bf k};n\tau) \;\;,
\end{equation}
with $g_{\alpha\beta}(\hat{\bf k};t)\equiv 
\langle\hat k_\lambda \sigma_{\alpha\lambda}(0)\vert
\hat k_{\lambda'} \sigma_{\beta\lambda'}(t)\rangle$, 
expresses the matrix of viscous transport coefficients, 
$\Lambda_{\alpha\beta}(\hat{\bf k})$, in
terms of a sum of time correlation functions of the reduced fluxes
\begin{equation}
\label{VIS2}  
\hat k_\lambda\sigma_{\alpha\lambda}(t)  \equiv 
\frac{1}{\tau}\sum_{j=1}^N\left(v_{j\alpha}(t)
{\bf \hat k}\cdot\Delta\bxi_j(t) + \Delta v_{j\alpha}(t)
{\bf \hat k}\cdot[\Delta\bxi_j^s(t)-{\bf z}^s_{jl}(t+\tau)/2] -
\frac{\tau \hat k_\alpha}{d}v_j^2(t)\right) , 
\end{equation}
where $\Delta\bxi_j\left(t\right) = \bxi_j\left(t+\tau\right) - \bxi_j\left(t\right)$, 
$\Delta v_{jx}(t)=v_{jx}\left(t+\tau\right)-v_{jx}(t)$
and $\Delta\bxi^s_j\left(t\right) = \bxi_j\left(t+\tau\right) - \bxi^s_j\left(t+\tau \right)$.
$\tau$ is the time step in the simulation and the prime on the sum in Eq. (\ref{VTCGK}) indicates that the $n=0$ term has the relative weight $1/2$. 
$\bxi_j(t)$ is the cell coordinate of particle $j$ at time $t=n\tau$,
while $\bxi_j^s$ is it's cell coordinate in the (stochastically) shifted
frame. ${\bf z}^s_{jl}(\tau)$ indexes pairs of cells which participate
in a collision event at time $\tau$; the second subscript, $l$, is the index 
of the collision vectors $\bsigma_l$ shown in Fig. \ref{fig.1}.
For example, for collisions characterized by $\bsigma_1$,
$z^s_{j1x}=1$ if $\xi^s_{jx}$ in Eq. (\ref{VIS2}) is one of the two cells on the
left of a supercell and $z^s_{j1x}=-1$ if $\xi^s_{jx}$ is on the right hand
side of a supercell; all other components of ${\bf z}^s_{j1}$ are zero.
In general~\cite{ihle_06_sdp}, the components of ${\bf z}^s_{jl}$ are either $0$, $1$, or $-1$.

Assuming only cubic symmetry, the most general form of 
$\Lambda_{\alpha\beta}(\hat{\bf k})$ in two dimensions is~\cite{ihle_05_ect}  
\begin{equation} 
\Lambda_{\alpha\beta}({\bf \hat k}) \equiv
\nu_1\delta_{\alpha\beta} + 
\nu_2\left(\delta_{\alpha\beta} - \hat k_\alpha \hat k_\beta \right)
+ \gamma \hat k_\alpha \hat k_\beta
+ \epsilon\ \hat k_\gamma \hat k_\rho I_{\alpha\beta\gamma\rho} , \label{lambdageneral}
\end{equation} 
where $\tensor I$ is the rank four unit tensor. 
$\nu_2$ is a new viscous transport coefficient associated with the 
non-symmetric part of the stress tensor, $\gamma$ is the bulk viscosity and $\epsilon$ is a viscosity 
coefficient which is nonzero only if there is cubic anisotropy.
In a simple liquid,
$\epsilon=0$ (because of invariance with respect to infinitesimal rotations),
$\nu=\nu_1$, and $\nu_2=0$ (because the stress tensor is symmetric in
$\partial_\alpha v_\beta$).

Since both particle streaming and the collisions contribute to momentum 
transport, there are two---kinetic and collisional---contributions 
to the transport coefficients. The kinetic contribution dominates 
at large mean free paths, $\lambda$, the collisional for $\lambda/a \ll1$. 
For the original SRD algorithm,
the kinetic contribution is isotropic, so that there is only one viscosity, 
$\nu$; the kinetic contribution to the bulk viscosity is also zero, as in a 
real ideal gas~\cite{tuzel_06_dcs}. However, because SRD collisions do not, in general, conserve angular momentum, the microscopic stress tensor is not symmetric and there 
is a collisional contribution\cite{ihle_05_ect} to $\nu_2$. 
It should be noted, however,  that 
the SRD algorithm can be modified to conserve angular momentum in two 
dimensions by introducing a unique, configuration dependent collision 
angle in each cell \cite{ryder_05_thesis}.   

In Refs. \cite{ihle_06_cpa,ihle_06_sdp} it was argued (for the nonideal 
model) that the 
probabilities of horizontal and vertical ($w$) and diagonal ($w_d$) collisions 
should be chosen so that the relaxation rate of the second moments of the 
velocity distribution function all decay at the same rate. This lead to 
the requirement that $w=1/4$ and $w_d=1/2$. Here we show that the same 
result follows from the requirement that the kinetic contribution to the 
viscous stress tensor is symmetric, so that there is only one viscosity, 
and $\epsilon=\gamma-\nu_2=0$. 
To see this, note first that for $\alpha=\beta=1$ with ${\bf \hat k} =\hat y$ and 
${\bf \hat k} = \hat x$, Eqs. (\ref{VTCGK})-(\ref{lambdageneral}) give
 $\gamma-\nu_2=-\epsilon/2=[\Lambda_{11}(\hat y)-\Lambda_{11}(\hat x)]$ and $\nu_1+\nu_2=\Lambda_{11}( \hat y)$. 
For large mean free paths, the term in parentheses in Eq. (\ref{VIS2}) reduces to 
$v_{j\alpha}(t){\bf \hat k}\cdot{\bf v}_j(t) - (\tau \hat k_\alpha/2)v_j^2(t)$ in two dimensions; using this result and making the assumption of 
molecular chaos, the sum in Eq. (\ref{VTCGK}) reduces to a 
geometric series in $g_{11}({\bf \hat k};\tau)$. 
The calculation of these quantities 
requires that the contributions from the horizontal, $g^H_{11}$, vertical, $g^V_{11}$, and diagonal 
collisions,  $g^D_{11}$, be calculated individually and then summed in the form 
\begin{equation} 
g_{11}({\bf \hat k};\tau) = w[g^H_{11}({\bf \hat k};\tau) + 
g^V_{11}({\bf \hat k};\tau)] 
+ w_dg^D_{11}({\bf \hat k};\tau) \;\;. 
\end{equation}
If $g_{11}(\hat y;\tau)\ne g_{11}(\hat x;\tau)$, 
there is a cubic anisotropy.  Ignoring fluctuations in the number 
of particles per cell, we find that  $g_{11}(\hat y;\tau) = 
g_{11}(\hat x;\tau)$ only if $w=1/4$ and $w_d=1/2$. In this case, the only non-zero viscosity is
\begin{equation} 
\nu\equiv\Lambda_{11}(\hat y) = \frac{\tau k_BT}{2}\left(
\frac{1}{A}\sqrt{\frac{2\pi}{k_BT}}[M_AM_B(M_A+M_B)]^{-1/2}-1\right)\;\;,
\end{equation} 
where  $M_A \equiv N_A / (L/a)^2 $ and 
$M_B \equiv N_B / (L/a)^2 $. In deep quenches, the density of the minority phase is very 
small, and the nonideal contribution to the viscosity approaches zero, i.e. 
$g(\tau) \sim 1 - O(\rho_A)$; in this case, the SRD collisions provide the 
dominant contribution to the viscosity. 

\section{Free Energy}

An analytic expression for the equation of state of this model can be derived 
by calculating the momentum transfer across a fixed surface, in much the 
same way as was done for the nonideal model in Ref. \cite{ihle_06_cpa}. 
The resulting expression for the nonideal contribution to pressure is
\begin{equation} 
\label{PTH} 
P_n = \left(w+\frac{w_d}{\sqrt{2}}\right)AM_AM_B \frac{k_BT}{a\tau} = 
\Gamma k_B T\rho_A\rho_B, 
\end{equation} 
where $\rho_A$ and $\rho_B$ are the densities of A and B and 
$\Gamma\equiv(w+w_d/\sqrt{2})a^3A/\tau$.    
In simulations, the total pressure can be measured by taking the ensemble 
average of the diagonal components of the microscopic stress tensor. In 
this way, the pressure can be measured locally, at the cell level. In 
particular, the pressure in a region consisting of $N_c$ cells is  
\begin{equation} 
\label{VSP} 
P = \frac{1}{\tau a^2 N_c}\left\langle\sum_{c=1}^{N_c}\sum_{i\in c} 
[{\tau v_{jx}^2 - \Delta v_{jx}z^s_{jlx}/2}]\right\rangle, 
\end{equation} 
where the second sum runs over the particles in cell $c$. 
The first term in Eq. (\ref{VSP}) is the ideal gas contribution to the pressure; 
the second term comes from the momentum transfer between cells involved in 
the collision indexed by ${\bf z}^s$. 
The results of measurements of the non-ideal contribution to the pressure 
obtained using Eq. (\ref{VSP}) are shown in Fig. \ref{fig.2} for $k_BT=0.006$ 
($\bullet$) and $k_BT=0.003$ ($\circ$). The lines are the theoretical 
predictions of Eq. (\ref{PTH}). For small values of $A$, the agreement between 
theory and simulation is excellent; deviations for larger $A$ are an indication that the model is no longer thermodynamically consistent~\cite{tuzel_06_ctc}. 

\begin{figure}
\twofigures[height=2.5in]{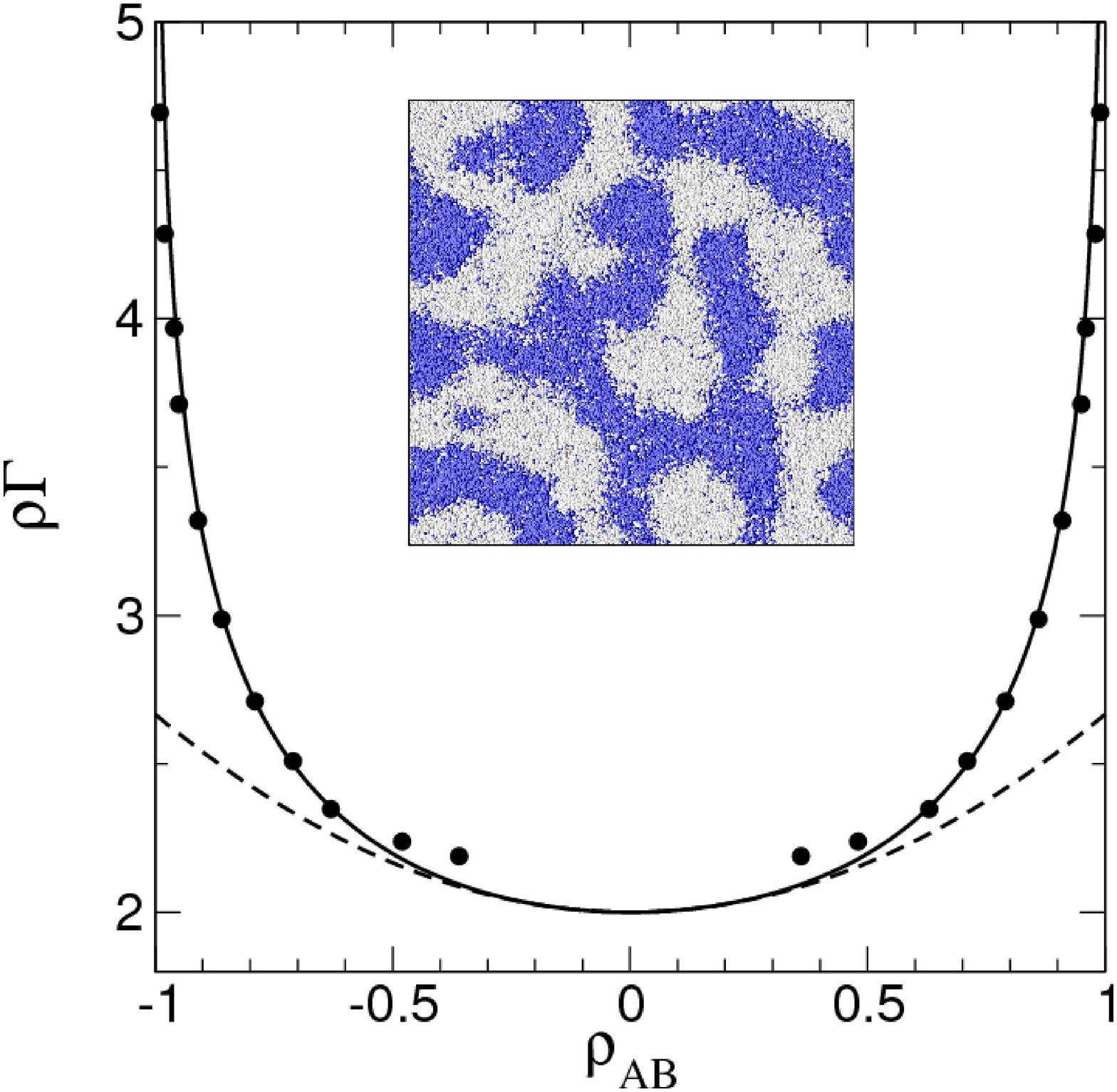}{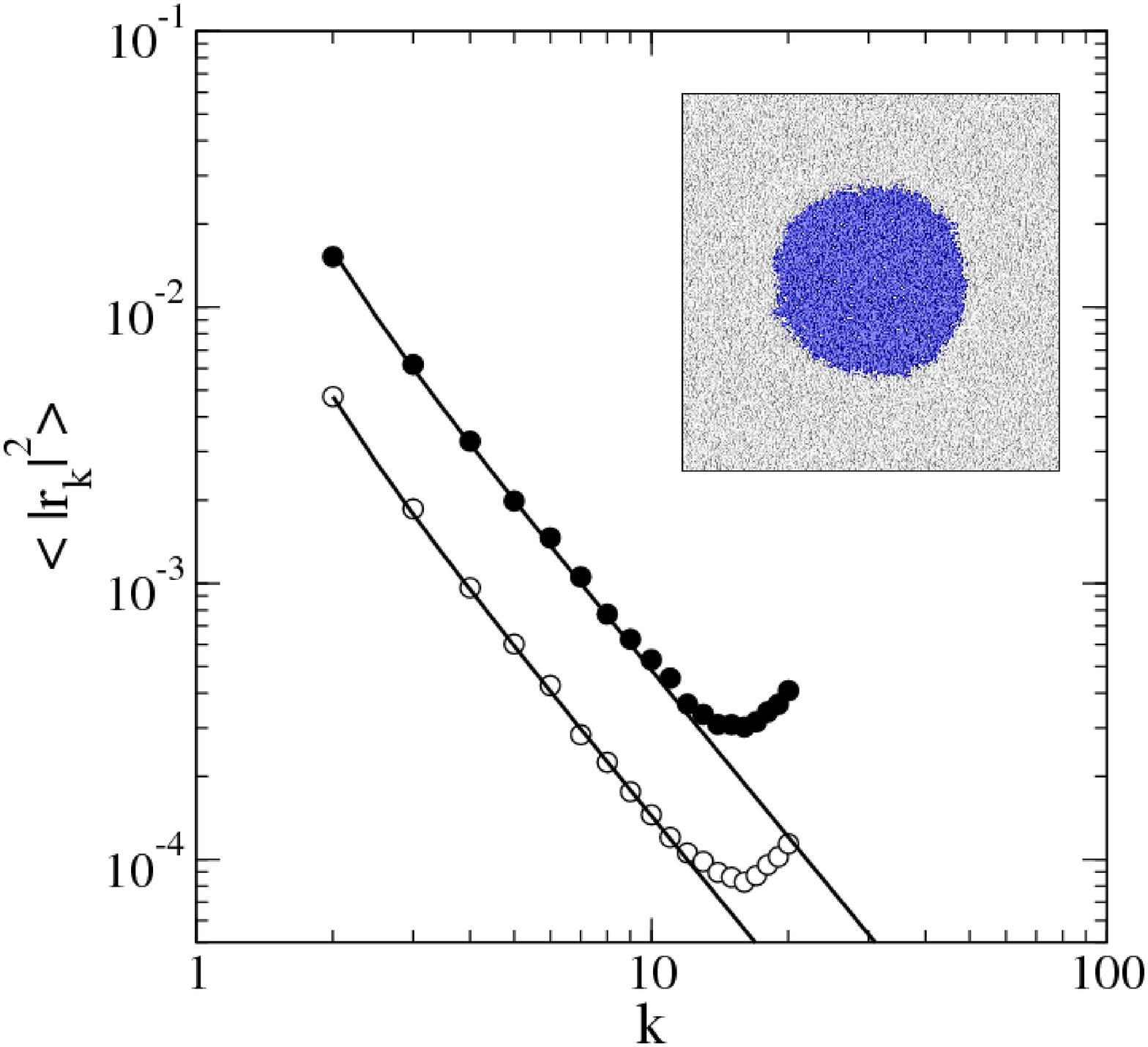}
\caption{Phase diagram of a $50\%$ A - $50\%$ B mixture. There is phase  
separation for $\rho\Gamma>2$. The inset shows a configuration $50,000$ time 
steps after a quench along $\rho_{AB}=0$ to $\rho\Gamma=3.62$. The dark (blue) and light (white) colored spheres are A and B particles, respectively. Parameters: $L/a=64$, $M_A=M_B=5$, 
$k_BT=0.0004$, $\tau=1$ and $a=1$.}  
\label{fig.3}
\caption{Dimensionless radial fluctuations, $\langle |r_k|^2 \rangle$, as a function of 
the mode number $k$ for $A=0.45$ ($\bullet$) and $A=0.60$ ($\circ$), 
with $k_BT=0.0004$. The average droplet radii are $r_0=11.95\,a$ and 
$r_0=15.21\,a$, respectively. The solid lines are fits to Eq. (\ref{ST}). The inset shows 
a typical droplet configuration for $A=0.60$ and $T=0.0004$. Parameters: $L/a=64$, $M_A=2$, $M_B=8$, $a=1$ and $\tau=1$.}  
\label{fig.4}
\end{figure}

Equation (\ref{PTH}) can be used to determine the entropy density, $s$. 
The ideal gas contribution to $s$ has the form \cite{callen_60_t}
\begin{equation}
\label{SI} 
s_{ideal} = \rho\,\varphi(T) - k_B[\rho_A\ln\rho_A + \rho_B\ln\rho_B], 
\end{equation} 
where $\rho=\rho_A+\rho_B$. Since $\varphi(T)$ is independent 
of $\rho_A$ and $\rho_B$, this term does not play a role in the current 
discussion. The nonideal contribution to the entropy density, $s_n$,  
can be obtained from Eq. (\ref{PTH}) using the thermodynamic relation~\cite{callen_60_t} 
\begin{equation} 
P_n/T =  s_n - \rho_A\partial s_n/\partial\rho_A - 
\rho_B\partial s_n/\partial\rho_B. 
\end{equation} 
The result is $s_n = - k_B \Gamma \rho_A\rho_B$, so that 
the total configurational contribution to the entropy density is 
\begin{equation}
\label{BIN6}
s = -k_B\left[\rho_A\ln\rho_A + \rho_B\ln\rho_B + 
\Gamma \rho_A \rho_B\right]. 
\end{equation}

Since there is no configurational contribution to the internal energy 
in this model, the mean field phase diagram can be determined 
by maximizing the entropy at fixed density $\rho$. The resulting 
demixing phase diagram as a function of $\rho_{AB}\equiv (\rho_A-\rho_B)/\rho$ 
is given by the solid line in Fig. \ref{fig.3}. The critical point is 
located at 
$\rho_{AB}=0$, $(\rho\Gamma)^*=2$. For $\rho\Gamma<2$, the order parameter 
$\rho_{AB}=0$; for $\rho\Gamma>2$, there is phase separation into 
coexisting A and B-rich phases. Simulation results for $\rho_{AB}$ obtained 
from density histograms are shown as bullets ($\bullet$). The dashed line is a plot 
of the leading singular behavior, $\rho_{AB}=\sqrt{3(\rho\Gamma-2)/2}$, 
of the order parameter at the critical point. As can be seen, the agreement 
between the mean field predictions and simulation are very good  
except close to the critical point, where the histogram method of determining 
the coexisting densities is unreliable and critical fluctuations 
could influence the shape of the coexistence curve.  

\section{Line Tension} 
 
A configuration after 50,000 time steps following a quench to point 
$\rho_{AB}=0$, $\rho\Gamma=3.62$ of the phase diagram is shown in the 
inset to Fig. \ref{fig.3}, and a snapshot of a fluctuating droplet 
at $\rho_{AB}=-0.6$, $\rho\Gamma=3.62$  is shown 
in the inset to  Fig. \ref{fig.4}. The amplitude of the capillary wave 
fluctuations of a droplet  is determined by the line tension, $\sigma$. 
In particular, for a droplet in an incompressible fluid, 
the mean square amplitude of fluctuations of the Fourier components of the 
(dimensionless) droplet radius, $\langle \vert r_k\vert^2\rangle$, is related to the line 
tension by the dispersion relation~\cite{tuzel_06_thesis} 
\begin{equation} 
\label{ST} 
\langle \vert r_k\vert^2\rangle = \frac{2k_BT}{\pi r_0\sigma}
\left(\frac{1}{k^2-1}\right), 
\end{equation} 
where $r_0$ is the mean radius of the droplet. 
Fig. \ref{fig.4} contains a plot of $\langle \vert r_k\vert^2\rangle$ 
as a function of mode number $k$ for $A=0.45$ ($\rho\Gamma=3.62$) and $A=0.60$ 
($\rho\Gamma=2.72$). Fits to the data yield  $\sigma a/k_BT\simeq2.9$ for $\rho\Gamma=3.62$ and $\sigma a/k_BT\simeq1.1$ for 
$\rho\Gamma=2.72$. Mechanical equilibrium requires that the pressure difference across 
the interface of a droplet satisfies the Laplace equation, 
\begin{equation}
\label{LE} 
\Delta p = p_{in}-p_{out} = \sigma/r_0. 
\end{equation} 
Using Eq. (\ref{VSP}), we have determined $\Delta p$ as a function of the 
droplet radius for $A=0.60$ and $k_BT=0.0005$. The results are plotted in the inset to 
Fig. \ref{fig.2}, where it can be seen that the Laplace relation is 
satisfied for the correct values of the line tension. 

\begin{figure}
\twofigures[height=2.41in]{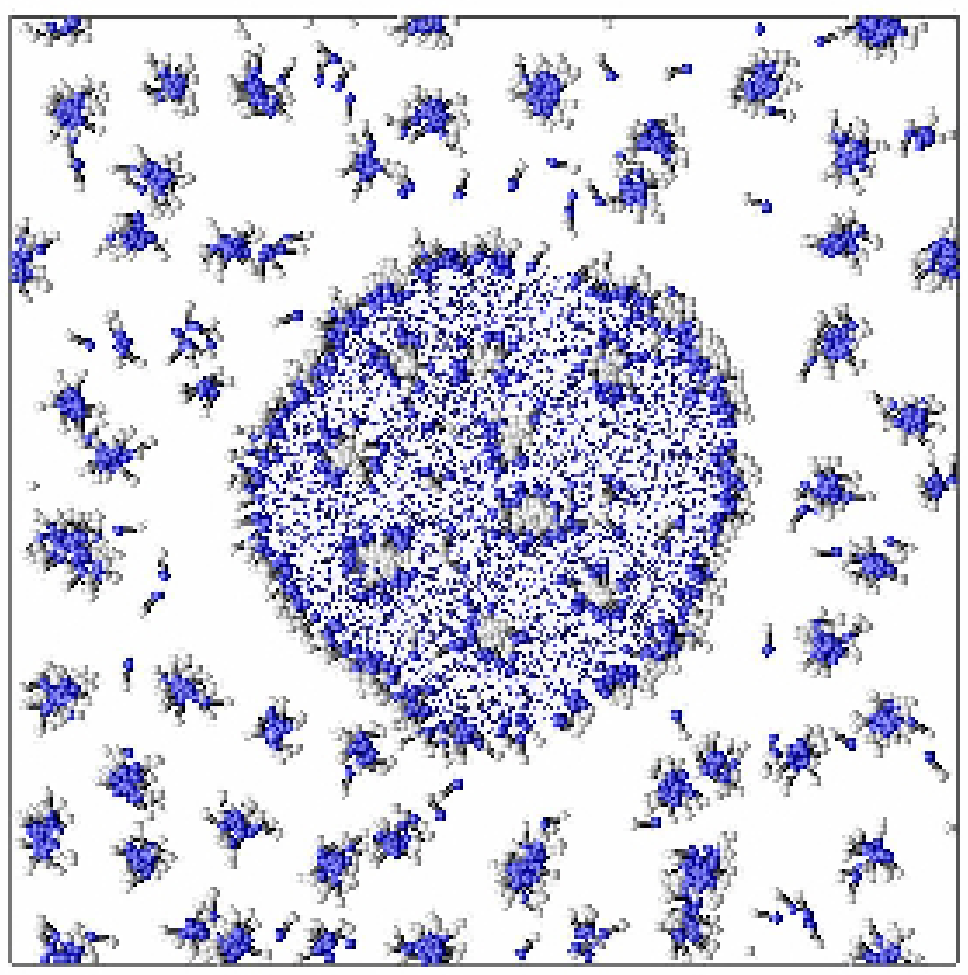}{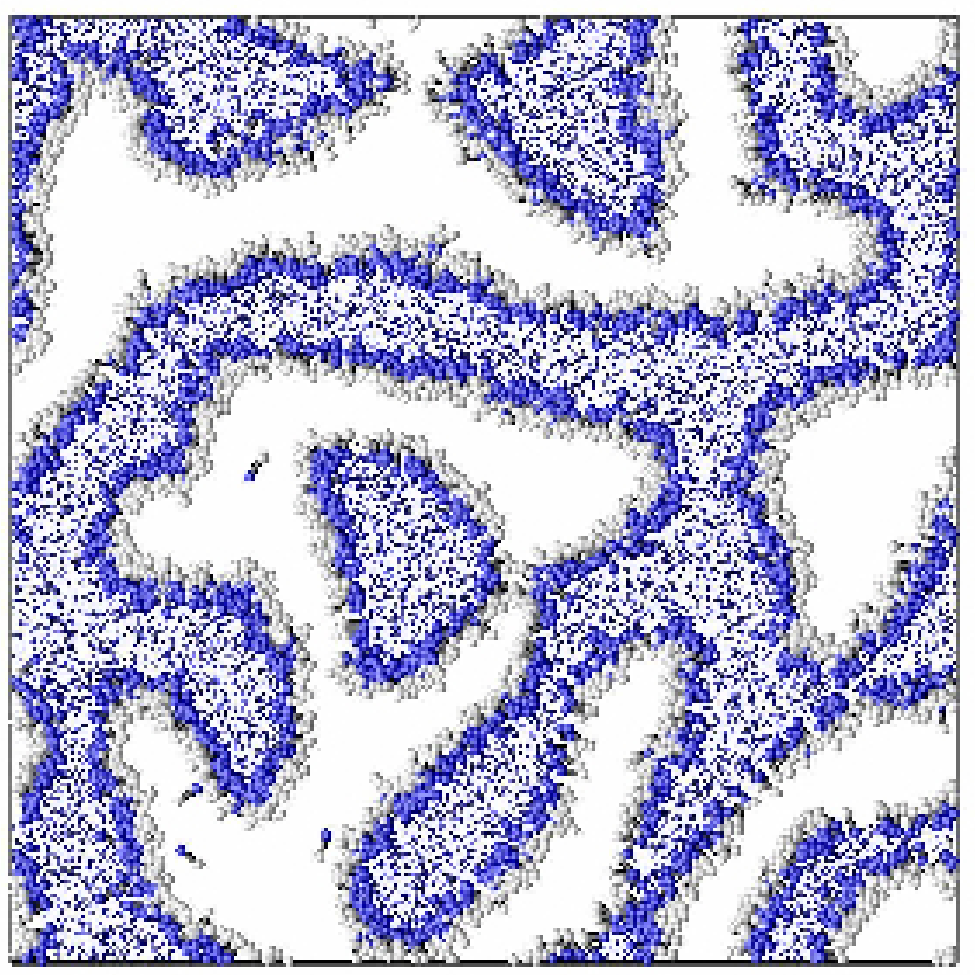}
\caption{A snapshot showing a droplet in a mixture with $N_A = 8,192$, $N_B = 32,768$ and $N_d=1,500$ dimers after $10^5$ time steps. The initial configuration is a droplet 
with a homogeneous distribution of dimers. The dark (blue) and light (white) colored spheres indicate the A and B particles, respectively. For clarity, A particles in the bulk are smaller, and B particles in the bulk are not shown. Parameters: $L/a=64$, $M_A=2$, $M_B=8$, $A=1.8$, $k_BT=0.0001$, $\tau=1$ and $a=1$.}  
\label{fig.5}
\caption{Typical configuration showing the bicontinuous phase for $N_A = N_B = 20,480$ and $N_d=3,000$. Parameters: $L/a=64$, $M_A=5$, $M_B=5$, $A=1.8$, $k_BT=0.0001$, $\tau=1$ and $a=1$.}  
\label{fig.6}
\end{figure}

The model therefore displays the correct thermodynamic 
behavior and interfacial fluctuations. It can also   
be extended to model amphiphilic mixtures by introducing dimers 
consisting of tethered A and B particles~\cite{tuzel_06_thesis}. If the A and B components of 
the dimers participate in the same collisions as the solvent, they 
behave like amphiphilic molecules in oil-water mixtures. The resulting model displays a rich phase behavior as a function of $\rho\Gamma$ and the number of dimers, $N_d$. 
We have observed both droplets and micelles, as 
shown in Fig. \ref{fig.5}, and a bicontinuous phase, as illustrated in 
Fig. \ref{fig.6}. The coarse-grained nature 
of the algorithm enables the study of large time scales with 
a feasible computational effort. 

\acknowledgments
Support from the National Science Foundation under Grant No. DMR-0513393 
and ND EPSCoR through NSF grant EPS-0132289 are gratefully acknowledged.


\end{document}